\documentclass[11pt,letterpaper]{article}
\pdfoutput=1

\usepackage{hyperref}
\usepackage{fullpage}
\usepackage{amssymb}
\usepackage{mathtools}
\usepackage{amsthm}
\usepackage{eucal}
\usepackage{dsfont}
\usepackage[T1]{fontenc}
\usepackage{lmodern}
\usepackage[stretch=10,shrink=10]{microtype}
\usepackage[capitalise]{cleveref}
\usepackage{color,soul}
\usepackage[linesnumbered,ruled,vlined]{algorithm2e}
\usepackage{graphicx}
\usepackage{framed}

\usepackage{amsmath}
\usepackage{amsfonts}

\allowdisplaybreaks[1]

\def\D{\CMcal{D}}
\def\E{\CMcal{E}}
\def\F{\CMcal{F}}

\def\H{\CMcal{H}}

\def\L{\CMcal{L}}

\def\N{\CMcal{N}}

\def\T{\CMcal{T}}

\theoremstyle{plain}
\newtheorem{theorem}{Theorem}[section]

\theoremstyle{definition}
\newtheorem{definition}[theorem]{Definition}

\newtheorem{remark}[theorem]{Remark}
\newtheorem{fact}[theorem]{Fact}

\newcommand {\bra} [1] {\ensuremath{ \left\langle #1 \right| }}
\newcommand {\ket} [1] {\ensuremath{ \left| #1 \right\rangle }}

\newcommand {\defeq} {\ensuremath{ \stackrel{\mathrm{def}}{=} }}

\newcommand {\Tr} {\ensuremath{ \mathrm{Tr} }}

\SetKwComment{Comment}{$\rhd\;$}{}
\SetKwInput{Input}{Input}
\SetKwInput{Output}{Output}
\SetKwInput{Promise}{Promise}

\newcommand {\email} [1] {\href{mailto:#1}{\texttt{#1}}}

\newcommand {\mytitle} {Quantum Insertion-Deletion Channels}

\newcommand{\suppress}[1]{}

\newcommand {\authorblock} [3] {
	\begin{minipage}[t]{0.3\linewidth}
		\centering
		{#1}\\[0.8ex]
		{\footnotesize {#2}\\[-0.7ex]
		\email{#3}}
	\end{minipage}\vspace{1ex}
}

\newcommand{\rI}{{\mathrm I}}
\newcommand{\tensor}{\otimes}
\newcommand{\Tra}[1]{\mathop{{\mathrm{Tr}}_{#1}} }
\newcommand{\kb}[2]{| #1\rangle\!\langle #2 |}

\hypersetup{
	pdfstartview={FitH},
	pdfdisplaydoctitle={true},
	breaklinks={true},
	bookmarksopen={true},
	bookmarksnumbered={false},
	pdftitle={\mytitle},
}

\newcommand {\Dave}   {Dave Touchette}
\newcommand {\Penghui} {Penghui Yao}

\newcommand {\CNO} {Department of Combinatorics and Optimization}
\newcommand {\IQCO} {Institute for Quantum Computing}
\newcommand {\UW} {University of Waterloo}
\newcommand {\PI} {Perimeter Institute for Theoretical Physics}
\newcommand{\SKL}{State Key Laboratory for Novel Software Technology, Nanjing University, Nanjing, 210023\\ P.R. China}

\begin{document}

\begin{titlepage}
\title{\textbf{\mytitle}\\[2ex]}

\author{
 \authorblock{Janet Leahy}{University of Calgary}{j.leahy886@gmail.com}
 \authorblock{\Dave}{\IQCO, \CNO, \UW, and \PI}{touchette.dave@gmail.com}
  \authorblock{\Penghui}{\SKL}{pyao@nju.edu.cn}
}

\clearpage\maketitle
\thispagestyle{empty}

\abstract{

We introduce a model of quantum insertion-deletion (insdel) channels. Insdel channels are meant to represent, for example, synchronization errors arising in data transmission. In the classical setting, they represent a strict generalization of the better-understood corruption error channels, and until recently, had mostly resisted effort toward a similar understanding as their corruption counterparts. They have received considerable attention in recent years. Very recently, Haeupler and Shahrasbi developed a framework, using what they call synchronisation strings, that allows one to turn insdel-type errors into corruption-type errors. These can then be handled by the use of standard error-correcting codes. We show that their framework can be extended to the quantum setting, providing a way to turn quantum insdel errors into quantum corruption errors, which can be handled with standard quantum error-correcting codes.

}

\end{titlepage}

\thispagestyle{empty}

\setcounter{page}{1}

\section{Introduction}

Quantum communication promises to offer a significant advantage over classical communication, both qualitatively and quantitatively.
It allows for unconditionally secure quantum key distribution~\cite{BB84}, a task impossible to achieve using only classical communication.
It also allows for the distribution of entanglement, which permits superclassical correlations.
Another notable example of a quantum advantage arises in the study of {\em communication complexity}, which investigates the minimum amount of communication required to a achieve a computational task in distributed computing. Quantum communication offers significant savings in numerous examples (see, e.g.,~\cite{Regev:2011} and the references therein).
Most often, these advantages are analysed in the setting of noiseless communication. To protect the quantum messages against the environment, which is probably the most essential to putting quantum computers and quantum communication into practice, numerous quantum error-correcting codes have been discovered. After decades of effort, correcting corruption errors, where some of the qubits are corrupted while the number of qubits is preserved, has become well-understood. However, the real world is more harsh. To execute reliable quantum communication in the real world, we must take {\em quantum synchronization errors}, which include both insertions and deletions, as well as other timing errors, into consideration. To the best of our knowledge, this hasn't been systematically studied yet.

Surprisingly, even designing error-correcting codes for classical synchronization errors turns out to be a very difficult task. Only in 1999 did Schulman and Zuckerman~\cite{SZuckerman:1999} present the first computationally efficient {\em insdel code} with constant rate, constant distance and constant alphabet size, resistant to insertions and deletions. Since then, a large body of work has been devoted to constructing better insdel codes in both one-way and interactive communication~\cite{GLi:2016, Guruswami:2017:DCH:3070907.3070977, Braverman2017coding,SWu:2017}. Recently, Haeupler and Shahrasbi~\cite{HShahrasbi:2017} introduced the novel notion of {\em synchronization strings}, which allow one to reduce insdel errors to corruption and erasure errors. Furthermore, these synchronization strings were shown in~\cite{Cheng2017SynchronizationSE, HaeuplerSTOC18p46} to admit a linear-time construction. Hence, their method provides an efficient black-box transformation from any error-correcting code into an almost-equally efficient insdel code.

In this paper, we investigate quantum insdel errors and successfully extend the synchronization string framework to the quantum setting. These results allow the transformation of quantum error-correcting codes into quantum insdel codes with almost equal efficiency.

\paragraph{Organization}
In the next section, we introduce preliminary notions. We then define a model of quantum insertion-deletion channels and provide a first coding scheme over a polynomial-size alphabet. In Section~\ref{sec:syncst}, we discuss the synchronization strings of Ref.~\cite{HShahrasbi:2017} and a means of extending that framework to the quantum setting for a large, constant-size alphabet. Finally, we discuss a coding scheme over qubit insdel channels.

\section{Preliminaries}

We briefly review the quantum formalism for finite-dimensional systems; for a more thorough treatment, we refer the interested reader to good introductions in a quantum information theoretical context \cite{NC00, Wat08, Wilde11}. For every quantum system $A$, we
associate a finite-dimensional Hilbert space, which by abuse of notation
we also denote by $A$. The state of quantum system $A$ is represented by
a {\em density operator} $\rho_A$, a positive semi-definite operator over the
Hilbert space $A$ with unit trace. We denote by $\D (A)$ the set of all
density operators representing states of system $A$. An important special case for quantum systems are {\em pure states}, whose
density operators have a special form:  rank-one projectors
$\kb{\psi}{\psi}$. In such a case, a more convenient notation is provided
by the pure state formalism: a state is represented by the unit vector
$\ket{\psi}$ (up to an irrelevant complex phase) upon which the density operator
projects.
 Composite quantum
systems are associated with the {\em (Kronecker) tensor product space} of the
underlying spaces, i.e., for systems $A$ and $B$, the allowed states of
the composite system $A \otimes B$ are (represented by) the density
operators in $\D(A \otimes B)$. We sometimes use the shorthand $AB$ for
$A \otimes B$.

Pure state evolution is represented by a
unitary operator $U_A$ acting on $\ket{\psi}_A$, denoted $U
\ket{\psi}_A$. Evolution of the $B$ register of a state $\ket{\psi}_{AB}$
under the action of a unitary $U_B$ is represented by $(\rI_{A} \otimes
U_{B})\ket{\psi}_{AB}$, for $\rI_{A}$ representing the identity operator
acting on the $A$ system, and is denoted by the shorthand $U_{B}
\ket{\psi}_{AB}$ for convenience. We occasionally drop the subscripts when the
systems are clear from the context.
In general, the evolution of a quantum system $A$ is represented by a
{\em completely positive, trace-preserving linear map} (CPTP map) $\N_A$ such
that if the state of the system was $\rho \in \D(A)$ before evolution
through $\N_A$, the state of the system is $\N_A (\rho) \in \D(A)$ after.
If the system $A$ is clear from the context, we might drop the subscript.
We
refer to such maps as {\em quantum channels}, and to the set of all channels
acting on $A$ as $\L(A)$.
We also consider quantum channels with different input and output
systems; the set of all quantum channels from a system $A$ to a system
$B$ is denoted $\L(A, B)$. Another important operation on a composite
system $A \otimes B$ is the {\em partial trace} $\Tr_B{[\rho_{AB}]}$ which
effectively derives the \emph{reduced\/} or marginal state of the $A$
subsystem from the quantum state $\rho_{AB}$.
Fixing an orthonormal basis $\{ \ket{i} \}$ for $B$, the
partial trace is given by~$\Tr_B{[\rho_{AB}]} = \sum_i (\rI \tensor \bra{i} )
\rho (\rI \tensor \ket{i})$, and this is a valid quantum channel in $\L(A\otimes B,
A)$.
Note that the action of $\Tra{B}$ is independent of the choice of basis chosen to represent it,  so we unambiguously write $\rho_A = \Tr_B{[\rho_{AB}]}$.


\section{Quantum insdel channels, and a first coding scheme}

\subsection{Definition of quantum insertion deletion channels}

Given a Hilbert space $\H$ of dimension $d$, define $\H_{\top}$ to be the Hilbert space of dimension $d+1$, such that $\H$ is embedded in the first $d$ dimensions, and any unit vector lying in the $(d+1)^{st}$ dimension is interpreted as the end-of-transmission symbol $\top$.

Similarly, given a Hilbert space $\H$ of dimension $d$, define $\H_{\bot}$ to be the Hilbert space of dimension $d+1$, such that $\H$ is embedded in the first $d$ dimensions, and any unit vector lying in the $(d+1)^{st}$ dimension is interpreted as the erasure symbol $\bot$.

We now define quantum insertion-deletion (insdel) channels. Notice that we must be careful and account for the fact that reading the content of a quantum register might corrupt its content, unlike with classical registers, which can be read without getting corrupted.

\begin{definition}\label{def:qinsdelchannel}[Quantum Insertion-Deletion Channel]
	Given $\delta \in (0, 1)$ and an input consisting of registers $A_1, A_2, ...A_n$, a \textbf{quantum insertion-deletion  (insdel) channel} over Hilbert space $\H$ producing (at most) a $\delta$-fraction of errors is defined by a noise function $\N\in \L ( \H^{\otimes n}, (\H_{\top})^{\otimes n+ n\delta})$ where there exist $p, q \in \mathbb{Z}^+$, $p + q \leq n\delta$, representing the number of deletions and insertions, respectively, such that $\N$ can be decomposed as:
	\begin{align*}
	\N = PAD \circ \N'.
	\end{align*}
	Here, $\N' \in \L ( \H^{\otimes n}, \H^{\otimes n-p+q})$ is a CPTP map with output registers $D_1,\ldots, D_{n-p+q}$ such that there exists a set of correctly transmitted indices, $S \subseteq \{1, 2, ... n\}$, with $|S| = n - p$, and an injective, strictly monotonic function $f: S \to \{1, 2, ... n -p +q \}$ for which $\N'$ transfers the contents of register $A_i$ to register $D_{f(i)}$, without otherwise affecting them, for all $i \in S$. In other words,
	\begin{align*}
	\N' = \F \circ \T,
	\end{align*}
	in which $\T$ is the operation described and $\F$ is an arbitrary CPTP map acting on the remaining registers, i.e. from $\{A_1, A_2, ..., A_n \} \setminus \{ A_i : i \in S \} $ to $\{D_1, D_2, ... D_{n -p +q}\} \setminus \{ D_i : i \in im(f) \}$.
	
	The map $PAD: \D(\H^{ \otimes n -p+q}) \to  \D((\H_{\top})^{\otimes n+ n\delta})$ simply acts by moving the contents of register $D_i$ to register $B_i$ for $i \in \{1, 2, ... (n-p+q) \}$ and inserts the end-of-transmission symbol $\top$ in registers $\{ B_{(n-p+q+1)},  ...,  B_{(n + n\delta)} \}$.
	
\end{definition}

\subsection{A solution for polynomial-size alphabets}

Suppose Alice and Bob have access to a quantum insertion-deletion channel over Hilbert space $\H_{ch}$, which they wish to use for one-way communication. Instead of having Alice transmit across the channel directly, we follow the channel simulation model~\cite{haeupler2017synchronization}, inserting intermediary entities $C_A$ and $C_B$. Alice and Bob can communicate directly with $C_A$ and $C_B$, respectively, but $C_A$ and $C_B$ must communicate over the quantum insertion-deletion channel. By having $C_A$ and $C_B$ implement an indexing scheme that effectively turns synchronization errors into corruptions or erasures, it appears from Alice's and Bob's perspectives that they are communicating across a regular quantum channel, with corruption-type noise, over a lower-dimensional space $\H_{sim}$. Thus, they can communicate reliably by using a sufficiently strong QECC whose codewords lie in $(\H_{sim})^{\otimes n}$.

To illustrate our idea, we present a simple coding strategy with a polynomial-size alphabet (i.e., one which can be represented by logarithmically many qubits). We assume for simplicity that Eve is not allowed to insert $\top$ symbols. Let $\sigma_{A_1 A_2... A_n} \in (\H_{sim})^{\otimes n}$ be a codeword that Alice wishes to send. The trivial indexing scheme involves $C_A$ adding a ``message number'' to each subsystem $A_i$ before sending it across the channel. This is done using the isometric encoding functions $\E_i \in \L (\H_{sim}, \H_{ch})$, whose actions on pure states are given by
$
\E_i(|a \rangle_{A_i}) \defeq |a \rangle |i \rangle,
$
in which $\dim(\H_{ch}) = \dim(\H_{sim}) \times n$.
$C_B$ measures the index registers, traces out the index systems, and arranges all the resulting registers in the order given by the measurement outcomes.
If an index does not appear, $C_B$ places an erasure flag in the $\H_{sim}$ register corresponding to that index.
Similarly, if an index appear twice or more, $C_B$ puts an erasure flag in the $\H_{sim}$ register corresponding to that index and discards the corresponding registers coming from the channel. This results in a state $\sigma'_{B_1 B_2 ... B_n} \in  \D(((\H_{sim})_\bot)^{\otimes n})$, which $C_B$ presents to Bob.

This strategy results in at most a $\delta$ fraction of errors, as can be seen by following what happens to the registers sent by Alice. Note that the content of  at least $n-p$ registers is transmitted correctly, due to the noise pattern,
which leaves the content unchanged apart from shifting the registers in unchanged order.
All of these registers will be restored by $C_B$, except at most $q$ of them. These correctly transmitted, unrestored registers will be due to insertion errors that caused the same index to appear more than once, resulting in all registers with the same index being erased. Thus, at least $n - p - q$ of the original registers are correctly restored by $C_B$.

Any corruption that occurs in the result must be due to the insertion of a $\H_{sim}$ register with an index corresponding to one of the deleted ones. Hence, it takes at least one of the $p$ ``deletions'' plus one of the $q$ ``insertions'' for each register corruption. Each of the $p$ deletions with no corresponding insertion for that index will be flagged as an erasure. Similarly, each of the $q$ insertions with no corresponding deletion for that index will result in an erasure flag. Denoting the number of corruptions by $c$, and the number of erasures by $e$, we get $2c + e \leq p + q \leq n \delta$. Counting an erasure as a half-error and a corruption as two half-errors (i.e. a full error), this means
 that at most $n\delta$ half-errors occur over the channel simulation. By using a QECC that can correct for this $\delta$ fraction of (half-)errors, Alice and Bob can then communicate reliably over the simulated channel.

However, we note that this strategy requires that $\dim(\H_{ch}) = \dim(\H_{sim}) \times n$.
This is reasonable, only leading to a constant decrease in communication rate, if $\dim(\H_{sim})$ is polynomial in $n$.
Otherwise, for constant-size $\dim(\H_{sim})$,
the communication rate drops to zero as the transmission length increases. To solve this issue, we can use the technique of synchronization strings, presented by Haeupler and Shahrasbi~\cite{HShahrasbi:2017}, which effectively allows one to index an arbitrary-length sequence of messages with a constant-size alphabet, with only a slight increase in the error rate.

\section{Synchronization strings and large, constant-size alphabets}
\label{sec:syncst}

\subsection{Synchronization strings and classical communication over insertion-deletion channels}

In the classical setting, a synchronization string $S$ of length $n$ refers to a string of characters from some alphabet $\Sigma_{syn}$ used to index a sequence of $n$ message characters.

Formally, the use of synchronization strings is captured in the following definition:

\begin{definition}~\cite{haeupler2017synchronization}[$(n, \delta)$-Indexing Algorithm]
	The pair $(S, \mathcal{D}_S)$ consisting of a synchronization string $S \in \Sigma^n$ and an algorithm $\mathcal{D}_S$ is a {\em $(n, \delta)$-indexing algorithm} over alphabet $\Sigma$ if for any set of $n \delta$ insertions and deletions represented by $\tau$ which alter $S$ to $S_\tau$, the algorithm $\mathcal{D}_S (S_\tau)$ outputs either $\bot$ or an index between $1$ and $n$ for every symbol in $S_\tau$.
\end{definition}

When sending a sequence of $n$ message characters across an insertion-deletion channel, $C_A$ attaches the $i^{th}$ character of $S$ to the $i^{th}$ message character, and sends the resulting pair across the channel. Upon receiving the message-character pairs from $C_A$, having been subject to potential insertions and deletions, $C_B$ applies the $(n, \delta)$-indexing algorithm to each of the indexing characters, which together form $S_\tau$. $C_B$ then outputs the message characters in the order corresponding to the indices returned by the algorithm. If multiple or no symbols in $S_\tau$ output an index $i$, the $i^{th}$ output position is filled with an erasure symbol. This process is referred to as the {\em indexing procedure}.

The quality of an $(n, \delta)$-indexing algorithm is determined by how accurately the indices returned by the algorithm match the original indices of the message characters they were paired with. When measuring this, we only care about correctly determining the indices of the message characters that were successfully tranmitted, i.e. those that were not inserted by the channel.

\begin{definition}~\cite{haeupler2017synchronization}[Misdecodings]
	Let $(S, \mathcal{D}_S)$ be an $(n, \delta)$-indexing algorithm. We say this algorithm has \textbf{at most $k$ misdecodings} if for any $\tau$ corresponding to at most $n \delta$ insertions and deletions, the number of successfully transmitted indices that are incorrectly decoded is at most $k$.
\end{definition}

Haeupler and Shahrasbi discuss a particular class of synchronization strings, called $\epsilon$-synchronization strings, that give rise to a good $(n, \delta)$-indexing algorithm. Furthermore, these strings can be made arbitrarily long, using an alphabet whose size only depends on the value of $\epsilon$.

\begin{fact}~\cite{haeupler2017synchronization}
	For any $\epsilon \in (0,1)$, $n \geq 1$, there exists an $\epsilon$-synchronization string of length $n$ over an alphabet of size $\Theta(1/ \epsilon^4)$.
\end{fact}

The decoding algorithm they present yields the following result:

\begin{fact}~\cite{haeupler2017synchronization}
\label{fact:ssdec}
	Any $\epsilon$-synchronization string of length $n$ along with the minimum relative suffix distance decoding algorithm form a solution to the $(n, \delta)$-indexing problem that guarantees $\frac{2}{1 - \epsilon} n\delta$ or less misdecodings.
	
	If $(S, \mathcal{D}_S)$ guarantees $k$ misdecodings for the $(n, \delta)$-indexing problem, then the indexing proceedure recovers the codeword sent up to $n\delta + 2k$ half-errors, i.e., the half-error distance of the sent codeword and the one received by the indexing procedure is at most $n\delta + 2k$.
\end{fact}

	\begin{remark}
This decoding algorithm is streaming and can be implemented so that it works in $O(n^4)$ time.
\end{remark}
It is worth noting that in this algorithm, $C_B$ decodes the index of the $i^{th}$ message using the sequence of all $i$ characters received so far, as opposed to just the most recently transmitted character. In this way, the encodings of the indices are really the prefixes of the synchronization string, rather than the individual characters themselves.

Thus, using the channel simulation model, Alice and Bob can communicate over an insertion-deletion channel, via $C_A$ and $C_B$, as if they were communicating over a regular corruption channel with a slightly higher error rate. If Alice has a message consisting of $m$ characters that she wishes to transmit across an insertion-deletion channel, she must first encode it using an ECC whose codewords are of some length $n$, and for which the minimum distance between codewords is larger than $n\delta + 2k$, where $k = \frac{2}{1-\epsilon}n \delta$ for some $\epsilon \in (0,1)$. If $C_A$ and $C_B$ follow the $(n, \delta)$-indexing procedure using an $\epsilon$-synchronization string of length $n$ and the algorithm given by Fact~\ref{fact:ssdec}, when Alice sends this codeword across the simulated channel, Bob will receive it with at most $n\delta + 2k$ half-errors. Given the strength of the ECC applied by Alice, Bob can then reliably recover the original message.

The key feature here is that the alphabet over which the synchronization string is constructed, $\Sigma_{syn}$, depends only on the value chosen for $\epsilon$, and does not depend on the length of the codewords transmitted. Thus, assuming that $| \Sigma_{ch} | \geq | \Sigma_{sim} | \times | \Sigma_{syn} |$, where $\Sigma_{ch}$ and $\Sigma_{sim}$ are the respective alphabets of the insertion-deletion and simulated channels, Alice can reliably send codewords of any length, provided that the ECC she uses is resistant to a $\delta (1 + \frac{4}{1-\epsilon})$ fraction of errors. In this way, the simulated channel gives only a constant slowdown in communication rate, regardless of the length of the transmission.
We show how to extend this to the quantum setting in the next section.
For binary channels, a bit more work is necessary, but again similar techniques to Ref.~\cite{haeupler2017synchronization} can be carefully extended to the quantum setting. We study binary quantum insdel channels in Section~\ref{sec:qubit}.

\subsection{Synchronization Strings in the Quantum Setting}
\label{sec:syncqu}

The techniques developed for $\epsilon$-synchronization strings can be almost directly translated into the quantum setting using the model presented in Section 3. Furthermore, using such $\epsilon$-synchronization strings solves the problem caused due to the communication rate decreasing with longer transmissions, which occurs when using the basic indexing scheme.

We define the {\em quantum indexing procedure} as follows:
Given an $\epsilon$-synchronization string $S \in (\Sigma_{syn})^n$ and a corresponding $(n, \delta)$-indexing algorithm, let $\sigma_{A_1A_2...A_n} \in (\H_{sim})^{\otimes n}$ be a quantum system over $n$ registers that Alice wishes to send across the simulated channel. Without loss of generality, we can assume that $\Sigma_{syn} = \{ 1, 2, 3, ... k \}$ for some $k$. Upon receiving $\sigma$, $C_A$ can then attach the $i^{th}$ character of the synchronization string to the $i^{th}$ register using the isometric encoding functions $\E_i \in \L ( \H_{sim}, \H_{ch})$, whose actions on pure states are given by
\begin{align*}
\E_i(|a \rangle_{A_i}) = |a \rangle |S_i \rangle,
\end{align*}
in which $S_i$ is the $i^{th}$ character of $S$ and $\dim(\H_{ch}) = \dim(\H_{sim}) \times k$.\\

The corresponding decoding procedure is almost exactly the same as described in Section 3. However, the measurement projectors applied by $C_B$ are given by $\{ \Pi_\ell \}$ for $1 \leq \ell \leq k$, where \begin{align*}
\Pi_\ell = \rI_{sim} \otimes | \ell \rangle\langle \ell|
\end{align*}
in which $\rI_{sim}$ is the identity on $\H_{sim}$. Then, instead of directly interpreting the outcomes as indices, let $S_\tau \in (\Sigma_{syn})^{n-p+q}$ be the string such that the $i^{th}$ character of $S_\tau$ is the measurement outcome produced in the $i^{th}$ register. The register indices are obtained by applying the $(n, \delta)$-indexing algorithm to the characters of $S_\tau$. These indices are used to produce the output system in the usual way: if an index was returned by the indexing algorithm exactly once, then the index system is traced out, and the resulting system is placed in the indicated output register. If an index was returned multiple or no times, then those registers are filled with the erasure symbol $\bot$. This results in a $n$-register quantum system in $\D(((\H_{sim})_\bot)^{\otimes n})$, which $C_B$ sends to Bob.

Since the processing of the measurement outcomes to produce register indices is entirely classical, we can directly apply results from Section 4. If $S$ is an $\epsilon$-synchronization string of length $n$, the $(n, \delta)$-indexing algorithm used by $C_B$ is the one given by Haeupler and Shahrasbi, and the ECC used by Alice to produce codewords of length $n$ is resilient to a $\delta(1 + \frac{4}{1-\epsilon})$ fraction of errors, the quantum indexing procedure results in reliable communication between the parties. As in the classical setting, $| \Sigma_{syn} | \in \Theta(1 / \epsilon^4)$, so the communication rate of the simulated channel depends only on $\epsilon$, and not on the length $n$ of the transmission. Thus, combining $\epsilon$-synchronization strings with the given quantum indexing procedure results in a simulated corruption channel with only a constant slowdown in communication rate.

\section{Qubit insertion-deletion channels}
\label{sec:qubit}

So far, we have handled channels with size growing with the size of the required synchronization string. What about constant-size input? We focus in this section on the worst-case, dimension-2 input; larger constant-size input can be handled similarly. The technique we use is similar to that used in Ref.~\cite{haeupler2017synchronization}: divide the message to be communicated into small, constant-size chunks (constant in the length of the message, but depending on the noise rate of the channel) of length $r$, and then transmit these chunks sequentially interspersed with header information. One important difference in the quantum setting is that we cannot ``read'' the symbols corresponding to the message, though we can read the header information as long as it is chosen independently of the content of the quantum message (apart from register ordering information, which we consider to be known information to the sender and separate from the possibly unknown quantum content in these registers). In fact, as long as we are careful about how we handle (potentially) quantum data, the same header as the protocol of Ref.~\cite{haeupler2017synchronization} works for us: first attach a ``barrier'' $10^s$ for some even parameter $s$ while insuring that the substring $0^{s}$ does not appear in the quantum data, and then attach a synchronisation string symbol (also free of the $0^{s}$ substring) to index this chunk. With this approach, we prove the following result.

\begin{theorem}
For a binary quantum insertion-deletion channel with error rate $\delta$ and an input quantum system on $n$ qubits, we can convert the insdel errors to $\Theta (n \sqrt{\delta \log (1 / \delta)})$ corrruption and erasure errors in a computationnally efficient way.
\end{theorem}

In fact, these $\Theta ( n \sqrt{\delta \log (1 / \delta)})$ errors have the following block structure: they are covered by $\Theta (n \delta )$ blocks of  $r = \Theta (\sqrt{\frac{\log 1/\delta}{\delta}})$ consecutively transmitted symbols.

\begin{proof}

We give an explicit protocol with the following parameters. We have chunks of size $r = \lceil \sqrt{\frac{\log 1/\delta}{\delta}} \rceil$, with a total of $N = \lceil n  \sqrt{\frac{\delta}{\log 1/\delta}} \rceil$ chunks. We also fix the barrier parameter to $s =2 \lceil c \log \frac{1}{\delta} \rceil$ for some constant $c$ that we specify later, and use an $\epsilon$-synchronization string $S$ of length $N$ on an alphabet $\Sigma_{sync}$ of size $2^l$ for some integer $l$. We consider input state $\sigma_{A_1 A_2 \ldots A_n }$, possibly unknown to Alice and correlated with some external system. Since we are hoping to turn insertion-deletion errors into corruptions and erasures, $\sigma$ should also be an encoding into some QECC if we are to correct these corruption errors.

We first need to remove any potential occurence of the substring $0^{s}$ in $\sigma$. We use a similar trick to Ref.~\cite{haeupler2017synchronization} and just embed the input system into a larger one in which we re-encode any potential $0^{s}$ substring into a different one. In more detail, look at any chunk of size $r$, and think of the corresponding bit string as a number in $[2^r]$. Further think of representing the corresponding number in base $2^{s/2} - 1$ and then map each of the symbols in this representation into strings of length $s/2$, excluding the $0^{s/2} $ string. In this way, there is one out of $2^{s/2}$ symbols which is lost, and we can then re-encode a string of length $r$ into one of length $r^\prime = \lceil r (1 + 2^{s/2}) \rceil$. Let $E_0$ be the isometry implementing this transformation from blocks of $r$ qubits to blocks of $r^{\prime}$ qubits. The protocol is then the following.\\

\begin{framed}
\textbf{Alice's encoding on input $\sigma_{A_1 A_2 \ldots A_n }$}
\begin{enumerate}
\item Split $\sigma$ into $N$ chunks of size $r$.
\item Apply $E_0$ to each of these chunks. 
\item Get $\sigma^\prime$ with N total chunks of size $r^\prime$, i.e. on $N r^\prime$ total qubits.
\item For $i = 1$ to $N$ do
	\begin{enumerate}
	\item send $10^s$ to Bob
	\item send $S(i)$ to Bob
	\item send next $r^\prime$ symbols of $\sigma^\prime$ to Bob
	\end{enumerate}
\end{enumerate}

\end{framed}

\begin{framed}
\textbf{Bob's decoding on channel output $\theta_{B_1 B_2 \ldots B_{N(r^\prime + l + s + 1)(1 + \delta)} }$}
\begin{enumerate}
\item For $i = 1$ to $N(1+\delta)$
	\begin{enumerate}
	\item If symbol is $\top$, get out of For loop
	\item Receive and measure incoming qubits until see substring $10^s$ on consecutive qubits
	\item Gather next $l$ qubits and put into next $E$ system, to later be interpreted as a symbol from synchronization string $S$.
	\item Gather next $r^\prime$ qubits, measure whether they lie in the support of the image of $E_0$
	\item If they do not, put state $\perp$ in next $D$ system.
	\item Else apply the inverse (on its image) of $E_0$  on these qubits, and put them in next $D$ system.
	\end{enumerate}
\item If less than $N(1+ \delta)$ $D \otimes E$ systems, put remaining ones in $\top$ state.
\item Perform decoding on the $N(1+\delta)$  $D \otimes E$ systems as in Section~\ref{sec:syncqu}.
\end{enumerate}

\end{framed}

We now argue that this protocol achieves the claim.
We first argue that the above protocol, before applying the decoding from Section~\ref{sec:syncqu}, effectively converts a qubit insdel channel into an insdel channel over a larger alphabet, of insertion-deletion rate $O(\delta)$.

First, notice that there are at most $\delta n$ $10^s$ header barriers that are changed and thus potentially lost.
This can happen by inserting a $1$ in the middle of the $0^s$ substring, or by deleting the leading $1$ or one of the $0$ in $0^s$, or even because a symbol was deleted from the previous potential $D \otimes E$ system and thus the first few bits of $10^s$ might become part of these systems.
This also potentially leads to the deletion of the subsequent $l$ synchronization string qubits and $r^\prime$ quantum data qubits, i.e. the deletion of the potential subsequent $D \otimes E$ system, but not further unless the following $10^s$ barrier is also lost.

Then, notice that there are at most $\delta n $ $10^s$ headers that are added. This can happen by deleting a $1$ in a substring of the form $10^k 10^l$ with $k+l \geq s$, or by inserting a substring $0^{k}$ into a $10^{l} 1$ substring, with $k + l > s$, or a combination; the important point is that at least one insertion or deletion is required per such error.

Finally, notice that at most $\delta n$ of the $l$ qubits corresponding to the synchronization string $S$ and a potential $E$ register, or of the $r^\prime$ qubits corresponding to the quantum data and a potential $D$ register, are changed (by insertion, deletion, or a combination).

Now, if a correct header is read and the corresponding subsequent potential $D \otimes E$ registers are also correct, then this counts as a correct transmission from the chunks of size $r$ qubits of $\sigma$ to the $D \otimes E$ registers from Section~\ref{sec:syncqu}, and the fact that this acts as a corresponding  insertion-deletion channel with $O(\delta)$ errors follows by the above bounds, since these are the only potential errors that can arise, for a total of at most $3 \delta n$.
 The result follows with appropriate choice of parameters, since we are working on chunks of size $r$ qubits.
\end{proof}

\section{Conclusion}

In this work, we introduced a model of quantum insertion-deletion channels, meant to represent synchronization errors arising in real-world communication. Our definitions carefully account for the fact that reading a quantum register disturbs its content. By extending the synchronization string framework of Haeupler and Sharhasbi, we were able to efficiently transform the problem of coding for quantum insdel channels to the more familiar task of coding for corruption error channels. We provide solutions for the harshest case, that of qubit insdel channels, after extending the framework for larger-size alphabet.

The definition we chose to adopt for quantum insdel channels remains simple to convey the main ideas of the framework, but the coding strategy could be adapted to handle more general insdel error models in the spirit of adversarial (one-way) corruption channels~\cite{leung2008communicating}.

It would be interesting to study two-way models of quantum insdel channels and to investigate how robust quantum communication complexity is to insdel-type errors. Much work has recently been done in the classical setting~\cite{Braverman2017coding, sherstov2017optimal, haeupler2017synchronization} toward understanding how interactive communication behaves under such insdel-type synchronization errors.

It would also be interesting to try to improve on the bounds obtained here. We focused on providing simple solutions to exhibit the power of the framework in the quantum setting, thus providing the first non-trivial bounds, but did not optimize parameters.

\section*{Acknowledgements}

We thank Debbie Leung for helpful discussions. Part of this work was done while J.L. held an URA award at the Institute for Quantum Computing (IQC), University of Waterloo, and was also supported in part by NSERC and CIFAR.  D.T. is supported in part by NSERC, CIFAR, Industry Canada as well as an NSERC PDF. IQC and PI are supported in part by the Government of Canada and the Province of Ontario. P.Y. is supported by the National Key R \& D Program of China under Grant 2018YFB1003202,  the China Youth 1000-Talent Grant and Anhui Initiative in Quantum Information Technology under Grant AHY150200. Part of this work was done while P.Y. visited PI.

\bibliographystyle{alpha}
\bibliography{references}

\end{document}